\documentclass{INTERSPEECH2023}

\interspeechcameraready

\usepackage{multirow}
\usepackage{xcolor}
\usepackage{cleveref}
\usepackage{lipsum}
\usepackage{blindtext}
\usepackage{bbding}

\title{Towards Selection of Text-to-speech Data to Augment ASR Training}
\name{Shuo Liu$^{1,2}$, Leda Sar{\i}$^1$, Chunyang Wu$^1$, Gil Keren$^1$, Yuan Shangguan$^1$, \\Jay Mahadeokar$^1$, Ozlem Kalinli$^1$}
\address{
  $^1$ Meta AI\\
  $^2$ University of Augsburg, Augsburg, Germany
}
\email{shuo.liu@uni-a.de, ledasari@meta.com}

\begin{document}

\maketitle
\begin{abstract}
This paper presents a method for selecting appropriate synthetic speech samples from a given large text-to-speech (TTS) dataset as supplementary  training data for an automatic speech recognition (ASR) model. 
We trained a neural network, which can be optimised using cross-entropy loss or Arcface loss, to measure the similarity of a  synthetic data to real speech. 
We found that incorporating synthetic samples with considerable dissimilarity to real speech, owing in part to lexical differences, into ASR training is crucial for boosting recognition performance. 
Experimental results on Librispeech test sets indicate that, in order to maintain the same speech recognition accuracy as when using all TTS data, our proposed solution can reduce the size of the TTS data down below its $30\,\%$, which is superior to several baseline methods.
%This paper explores methods for selecting appropriate synthetic speech samples as supplementary data for training an automatic speech recognition (ASR) model. 
%Given a large text-to-speech (TTS) dataset, we aim to use only a subset of it while maintaining the same speech recognition accuracy as when using all data. 
%To determine which kinds of TTS samples, those more similar or dissimilar to the real speech, are more useful in augmenting ASR training, we suggest to train a model to quantify the agreement between a TTS data and real speech.  
%Our experiments reveal that synthetic samples that are of considerable dissimilarity from real speech, partly due to lexical discrepancies, contain crucial information for boosting ASR training. 
%Testing results on Librispeech evaluation sets indicate that our proposed method is superior to several baseline methods, reducing the TTS data size down below its $30\,\%$ while preserving the ASR performance as the model trained using all available TTS data.
\end{abstract}
\noindent\textbf{Index Terms}: Data augmentation, text-to-speech, automatic speech recognition, synthetic speech, Transducer

\let\thefootnote\relax\footnotetext{{}\\[-4mm]\indent\ Work was done when Shuo was interning at Meta AI.}

\section{Introduction}
% TTS for data augmentation
Speech synthesis, or Text-to-Speech (TTS), is a fast developing technology that generates speech from given text. 
It has reached the stage that it can create audio that closely resembles natural human voice, allowing for the use of synthetic data to improve the training of Automatic Speech Recognition (ASR) models \cite{rosenberg2019speech,wang2020improving,ueno2021data}. 
%Recent works have found that exploiting synthetic speech data generated, for example, by using a text-to-speech (TTS) model, as additional training data is an effective data augmentation method that improves the performance of automatic speech recognition (ASR) \cite{rosenberg2019speech,wang2020improving,ueno2021data}. Ideally  
% TTS data size can be very large.
Ideally, creating more synthetic data for ASR training has always the potential to improve the accuracy of speech recognition. 
However, as the quantity of the training data grows, more training time and computational resources are demanded.
% redundancy in TTS data
On the other hand, since TTS speech data are often produced with limited alternative constrains in terms of, for instance, speakers, speech speed and pitch, the resultant synthetic dataset may include a substantial amount of redundancy. 
Hence, it should be possible to minimise the data size by only choosing more typical and representative samples in a given large TTS dataset to achieve a more cost-efficient ASR training.

% Existing data selection for ASR
Recent research \cite{hu2022synt++,lu2022unsupervised} have demonstrated some data selection strategies that are effective for this data reduction purpose. 
In \cite{hu2022synt++}, it is suggested to exploit rejection sampling  to accept or reject synthetic samples such that the distribution of the selected synthetic data is close to that of the real speech.   
The distribution is represented in five dimensions depending on the outputs of a pre-trained ASR model, such as Xent loss, CTC loss, and token lengths, etc. 
Thus, the chosen synthetic data should own high similarity to actual speech.
In \cite{lu2022unsupervised}, the objective is to choose data from a general pool that reflects a domain of interest, in order to build a domain-optimised ASR model.
For this purpose, two language models are separately trained on the data from the target domain and the general domain using discrete speech tokens as input. A domain relevance score is computed based on the output probabilities from the target and general domain language models. The general domain samples scored highest are selected for training the domain specific ASR model.
% Existing data selection for NLP
Data selection is also considered essential for training natural language processing (NLP) models efficiently \cite{settles2008analysis,hazra2019active}. To empower an NLP model with more lexical information, however, it has been advised to choose the data of low certainty for training so that extra linguistic knowledge may be included.

%On the other hand, data selection works in the NLP domain has presented the benefit of choosing data of lowest certainty, because these data can induce more additional knowledge in the training of the model \cite{settles2008analysis,hazra2019active}.  

% Similarity vs dissimilarity, agreement vs disagreement
When selecting TTS data for ASR training, a question remains unanswered is whether we should choose the data that are similar to or distinct from the original real speech. From the acoustic perspective, we may want to select the samples of high similarity to ensure signal quality. However, these data may contain much less additional information and hence can only marginally improve an ASR model; On the other hand, a TTS sample that differs significantly from the original real speech may indicate the synthetic audio is contaminated with noise or artefacts produced during the synthesis process, which can be detrimental to the ASR training.
%Facing to the same situation, when selecting TTS data for ASR training, it has not been explored whether their similarity or dissimilarity from the real speech is more important. Intuitively, too similar samples may contain less additional information, and thus give limited contributions to ASR improvement; while the TTS samples that are too different from real speech may be due to the contamination from noise and artifacts introduced by TTS model.
% our methods: LM based method from Hubert representation, bin classifier methods
% we present our method to find suitable TTS data for the training of ASR, aiming to reduce the data size while preserving the ASR performance. Our method employs a simple neural network, a Gated Recurrent Unit (GRU), to learn the embedding of an utterance, and compare the embedding of the synthetic sample to that of the real speech. We analyse the efficacy of TTS data of different agreement levels to real speech. By doing this, we finally reduce the data needed for ASR training. 
To choose the appropriate TTS data for ASR training, our methodology compromises the similarity and discrepancy of the synthetic and the original real speech. 
In particular, we use a basic recurrent neural network, a Gated Recurrent Unit (GRU), to generate speech embedding. Then, we compare the embeddings of the real and synthetic speech samples and use two scoring methods to reflect their agreement. For both scoring methods, when selecting the TTS data inside a certain scoring range,  we can lower the needed TTS data size while preserving the same level of ASR performance as using all available TTS data.

% paper outline.
%The rest of the paper is organised as follows. First, we explain our proposed method in \Cref{sec:method}, followed by the description of our experiments in \Cref{sec:exp}. Before the conclusion, we will analyse and discuss the effectiveness of our solution in \ref{sec:discuss}.
%\section{Related Work}
% classic data selection method - baseline methods. Inspired by these methods, we used LM scores...
% The method of binary classifier -> anti-proof/ fake audio detection

\begin{figure*}[t]
\hspace*{0.6cm}
  \centering
  \includegraphics[scale=0.4]{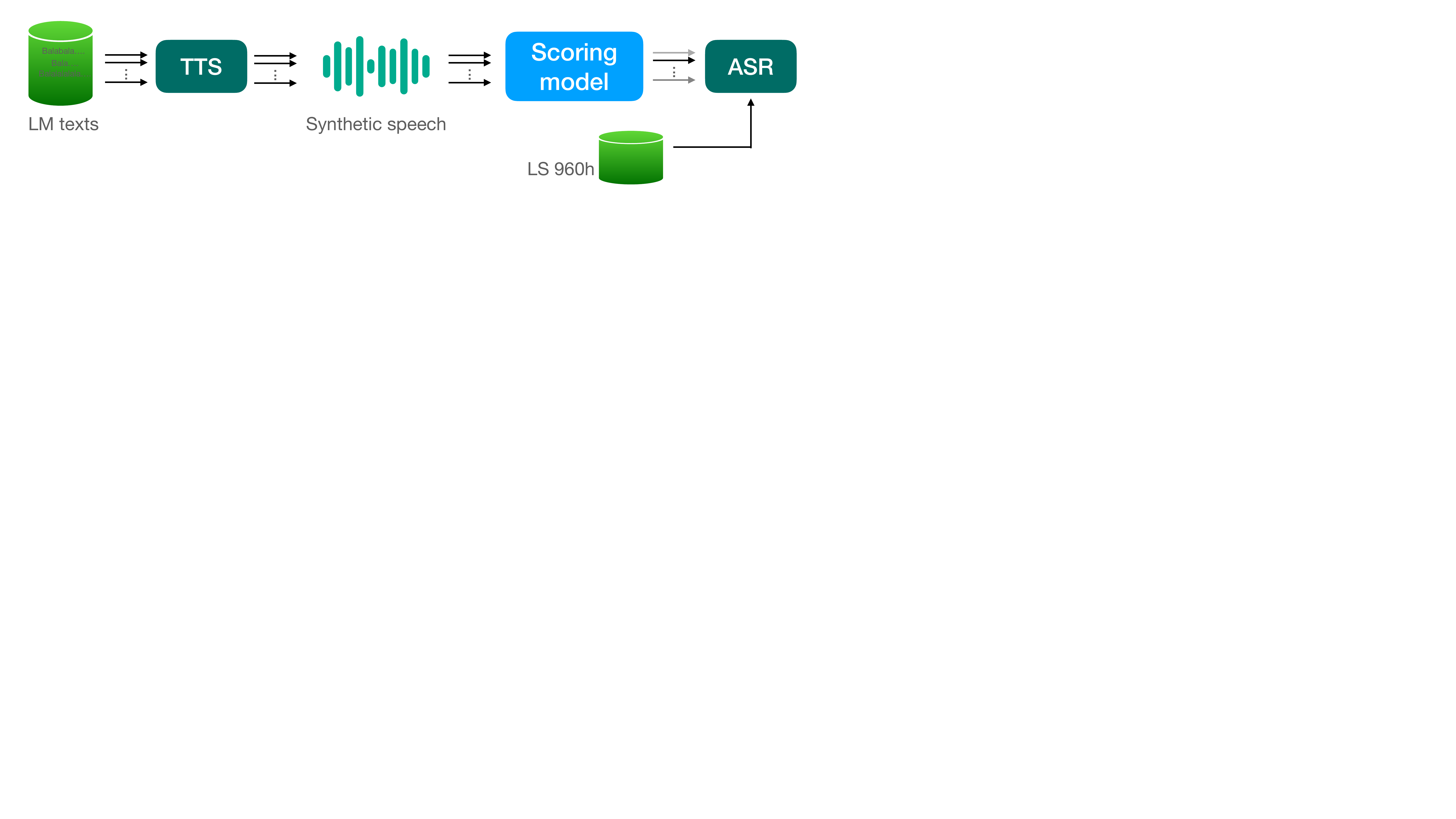}
  \vspace{-12.6cm}
  \caption{System overview}. %\textcolor{red}{LM texts $->$ additional text resources, LS 960h $->$ original real dataset}}
  \label{fig:performance1}
  \vspace{-0.8cm}
\end{figure*}

% THIS LIST IS PROVISIONAL - pending final version from Kate.
% \setlist{noitemsep,topsep=0pt,parsep=2pt,partopsep=0pt,leftmargin=1em}

\section{Methodology}
\label{sec:method}
% Please give an overview of the system by referring to Fig. 1.
% Something like: 
%As shown in Fig. \ref{fig:performance1}, we start from additional text resources, use a TTS model to generate synthetic speech files. We pass these files through a scoring model to filter out some files. The remaining files are then combined with the original real dataset to train an ASR model. For the test phase, we use only real test data.
%In this section, we present two data selection methods to choose useful synthetic samples produced by the TTS system. These methods are: (1) an acoustic unit LM score based approach in \Cref{ssec:lms}, and (2) a binary classifier based approach that uses the similarity/difference between real and TTS speech data in \Cref{ssec:bc}. Since our focus is on data selection, the details of the ASR and TTS components will be presented in the Experiments section (\Cref{ssec:asrtts}). 
The system overview used in this work is given in \Cref{fig:performance1}. We generate synthetic speech files from additional text resources using a TTS model. A scoring model is then trained to select some of these audio files that facilitate ASR training. These files are then added to the original real dataset to train an ASR model. For the test phase, we use only real test data.
%
%In this following, we describe our method to develop the scoring model.  
%The details of the ASR and TTS components will be presented in \Cref{ssec:asrtts}. 

%\subsection{Binary Classification Based Scoring}
%\label{ssec:bc}
%\vspace{-0.2cm}
To develop the scoring model, we train a two-layer Gated Recurrent Unit (GRU) model to extract speech representation from an input audio file. The output of the last time-step is fed into a fully-connected (FC) layer, which is then optimised using binary cross-entropy (BCE) or Arcface loss \cite{8953658}. When using the BCE loss, the FC output is linearly projected to the two classes, real or synthetic speech, through an additional FC layer. Arcface loss adds a margin parameter to the angle between the normalised speech representation and the weights of the FC layer that is implemented in the loss function. The insertion of the margin tends to increase the separability between speech representations of distinct classes.    
%We train a two-layer Gated Recurrent Unit (GRU) model to extract speech representation from an input sample, for which the embedding output of the last time-step is fed into a fully-connected layer and then optimised using a cross-entropy or the Arcface loss [ref?]. 
%After training the model using cross-entropy loss, the Softmax output of the predicted logit is used to score an utterance. We take the score of $0.5$ as the boundary to distinguish the real and TTS data, i.e., when the score is above $0.5$, the input utterance is recognised as real data, otherwise the TTS data. Intuitively, when a TTS sample is scored closer to $1$, it is recognised more like real speech, hence it is seen as having high similarity to real speech. A dissimilar TTS sample, however, may indicate more additional information that are helpful in training an ASR model. Thus, it is challenging but necessary to find a proper scoring range to filter out the suitable TTS data. 

Real speech is regarded as class 1 and synthetic speech as class 0 in the training of the GRU model using BCE loss. 
After training,  we calculate the Softmax value of the output logit to score an input audio, and the score of 0.5 can be seen as the threshold for distinguishing between a real or synthetic speech. 
If the score exceeds $0.5$, the input is recognised as a real data, otherwise it is classified as a TTS data. 
Hence, when a TTS sample's score is closer to $1$, it is more likely to be identified as real data, indicating its higher similarity to real speech. 
However, as mentioned earlier, a dissimilar TTS sample may provide extra information needed to enhance ASR training. Therefore, it is necessary, however challenging,  to find a proper scoring range for selecting the appropriate TTS data. 

%To measure the similarity and difference between a real and TTS speech data, a more straight forward solution is to compute the cosine similarity of the learned speech embeddings. We apply Arcface loss to GRU embedding of the last time-step, aiming to improve the discriminativeness of the representations of the two classes. For data selection, we compute the average speech embedding of the real utterances, and then compute its cosine similarity to a TTS sample. Similarly, low similarity can indicate more additional information that can contribute to the ASR training. 

A more straightforward approach to quantify the similarity between real and synthetic speech is to compute the cosine similarity of their speech representations. In an attempt to improve the representation separability, Arcface loss is applied to the GRU model's optimisation. Based on the real samples in training set, we next calculate the average embedding for real speech, named as average real embedding. To score a TTS sample, the cosine similarity between its embedding and the average real embedding is computed.
Again, the TTS samples scored with low similarity to real speech may signal the presence of new information that has the chance to contribute to the ASR training.
Notably, since our proposed method exploits only a simple neural network, it allows the rapid scoring of synthetic speech samples compared to previous work that typically relies on a pre-trained ASR model, particularly when the size of the involved TTS dataset is enormous.

\section{Experiments}
\label{sec:exp}
\vspace{-0.1cm}
\subsection{Dataset}
Our scoring models are trained based on the LibriSpeech corpus \cite{panayotov2015librispeech}. It consists of around 960\,hours of read, clean speech derived from over 8\,000 public domain audiobooks and has its own train, development, and test splits. We apply the TTS model described in \Cref{ssec:asrtts} to the Librispeech transcriptions to generate the synthetic audio files. During training, the performance of the scoring models are monitored on the Librispeech development set without touching the test set.
%As additional text resources for generating the synthetic speech files for ASR training, we employ $10\,\%$ of data from language model (LM) corpus, resulting in a total of 4.11 million utterances.  % + introduction of LM data.  
As additional text resources to generate the synthetic speech files for ASR training, we employ $10\,\%$ of data from the language model (LM) corpus provided in Librispeech corpus, yielding a total of 4.11 million utterances. 

\subsection{TTS \& ASR Models}
\label{ssec:asrtts}
%\subsubsection{TTS model}
%\label{ssec:tts}
The TTS model has a multi-stage framework. It begins with a rule-based Grapheme-to-Phoneme (G2P) Conversion module that transforms the input text into various levels of linguistic representations. These features are consumed by a prosody model \cite{9414809} and a spectrum model \cite{wu21b_interspeech} to generate spectrum features. The prosody model consists of multiple Transformer layers \cite{vaswani2017attention}, with each layer contains a multi-head attention module and a feed-forward layer. In addition, residual connection and layer normalisation are performed to each transformer layer. Finally, a neural vocoder using WaveRNN achitecture \cite{pmlr-v80-kalchbrenner18a} is used to transform the features into audio waveform. 

%\subsubsection{ASR model}
%\label{ssec:asr}
The ASR model used in this work processes a log Mel-spectrogram with $80$ Mel bands, the spectrogram is created by taking the Short-time Fourier transfrom (STFT) of a speech signal with a window size of $25$ms and hop size of $10$ms. 
%Each speech sample lasts between $0.5$ and $10$. 
SpecAugment is applied to strengthen the robustness of the ASR model \cite{park2019specaugment}. The model has the output vocabulary of $5000$ sentence pieces~\cite{kudo2018sentencepiece} estimated over the $960$-hours Librispeech training set~\cite{panayotov2015librispeech} (LS 960h).  
The Mel-spectrogram is linearly projected to $128$ channels using an encoder, and every four time steps are concatenated along the feature dimensionality to squeeze the total frames. Then a stack of $20$ efficient memory transformer (Emformer~\cite{shi2021emformer}) layers are used to capture the temporal context.
%The mel-spectrogram is linearly projected to $128$ channels using an encoder, and every four time steps are concatenated along the feature dimensionality, resulting in the reduction of total frames. Then a stack of $20$ efficient memory transformer (Emformer~\cite{shi2021emformer}) layers are used to capture the temporal context, each of the Emformer layer has $8$ attention heads, a feed-forward module with the dimension of $2048$ and GELU activation~\cite{hendrycks2016gaussian}, and a layer normalisation layer \cite{ba2016layer}. The Emformer output is then fed into a recurrent neural transducer (RNN-T). Using a 3-layer LSTM model ~\cite{hochreiter1997long}, the Emformer predictor generates embeddings of the size of $512$ based on all the previous predicted symbols. The outputs of the Emformer encoder and predictor are combined and projected to a probability distribution over a vocabulary. 
The Emformer output is then fed into a recurrent neural transducer (RNN-T). Using a 3-layer LSTM model ~\cite{hochreiter1997long}, the Emformer predictor generates embeddings of the size of $512$ based on all the previous predicted symbols. The outputs of the Emformer encoder and predictor are combined and projected to a probability distribution over a vocabulary.
%The output of each recurrent layer was regularized by means of dropout~($p=0.3$) and lastly projected to \qty{1024}{} layer-normalized channels. The additive combination of encoder and predictor outputs was projected to a probability distribution over the output vocabulary plus a blank token.

\begin{table*}[ht!]
  \begin{center}
  \caption{Testing results, WER [$\%$], of different data selection approaches on Librispeech test set. \textbf{\# add. utter} indicates the number of TTS samples selected to be added to Librispeech 960h (LS) training data. The proportion of utterances selected from the synthetic samples generated based on LM text database are additionally given, where ``high'' and ``low'' denote the TTS samples are selected from those with highest or lowest scores, respectively. The binary classifiers are trained using Cross-entropy (Xent) or Arcface loss.}
  \vspace{-0.1cm}
  \begin{tabular}{l | l | c c c c }
    \toprule
    \textbf{Method} & \textbf{\# add. utter. [M]} & \textbf{dev-clean} & \textbf{test-clean} & \textbf{dev-other} & \textbf{test-other}\\
    \midrule
    \textbf{LS} & $-$ & $3.2$ & $3.5$ & $9.2$ & $8.7$\\
    \textbf{LS + all synthetic speech} & $4.11$ ($100\,\%$) & $2.67$ & $2.97$ & $8.39$ & $8.05$\\
    \hline
    \textbf{Random} & $1.23$ ($30\,\%$) & $2.87$ & $3.21$ & $8.63$ & $8.63$\\
    \hline
    \multirow{2}{*}{\textbf{Confidence score}} & $1.23$ ($30\,\%$ high) & $3.10$ & $3.24$ & $8.90$ & $8.65$\\
     & $1.23$ ($30\,\%$ low) & $2.77$ & $3.05$ & $8.46$ & $8.28$\\
    \hline
    %\multirow{2}{*}{\textbf{Sentence entropy}} & $1.23$ ($30\,\%$ high) & $2.86$ & $3.11$ & $8.82$ & $8.27$\\
    %& $1.23$ ($30\,\%$ low) & $2.97$ & $3.11$ & $8.78$ & $8.37$\\
    %\hline
    \multirow{2}{*}{\textbf{ULM accuracy}} & $1.50$ ($36\,\%$ high) & $2.72$ & $2.97$ & $8.63$ & $8.22$\\
     & $1.23$ ($30\,\%$ low) & $2.91$ & $3.13$ & $8.58$ & $8.31$\\
    \hline
    \multirow{2}{*}{\textbf{ULM perplexity}} & $1.50$ ($36\,\%$ high) & $2.87$ & $2.99$ & $8.96$ & $8.35$\\
    & $1.23$ ($30\,\%$ low) & $2.76$ & $3.01$ & $8.39$ & $8.20$\\
    \hline
    \textbf{Binary classifier (Xent)} & $1.10$ ($27\,\%$) & $2.78$ & $2.96$ & $8.42$ & $8.16$\\
    \textbf{Binary classifier (Arcface)} & $1.23$ ($30\,\%$) & $2.81$ & $2.97$ & $8.46$ & $8.09$\\
    \bottomrule
  \end{tabular}
  \label{tab:result}
  \end{center}
  \vspace{-0.7cm}
\end{table*}

\subsection{Settings}
For a fair comparison of different data selection methods, the training parameters for all the ASR models developed throughout this work are consistent. The models are optimised by minimising the transducer loss using an Adam optimiser with an initial learning rate of $0.001$. The total number of training epochs is $200$, and beginning with the $60$ epoch, the learning rate annealing is applied at a shrink rate of $0.96$ per epoch.   
The maximum number of tokens in a training batch is capped to $40000$ with the number of utterances not exceeding $1000$.
The ASR models are trained using $32$ A100 GPU units.

Both layers of our GRU scoring model have hidden units of $256$.  
The dimensionality of the subsequent fully-connected layer is set to $64$ when the model is trained with BCE loss.  Its optimisation employs a batch size of $64$ and Adam optimiser with a fixed learning rate of $0.0001$.  When using Arcface loss for training, the scale and margin parameters are set to $10$ and $0.5$, respectively. A lower learning rate of $0.00001$ is applied to ensure stable convergence.

\subsection{Baseline Methods}
\vspace{-0.1cm}
As the primary comparison result, we first evaluate the effect of adding all available TTS data to the original Librispeech training set as the new training data.
In addition, we compare our data selection approach with three baseline methods: random selection, a method depending on confidence score, and an acoustic unit language model-based scoring method.
%We consider three baseline methods in addition to the case of not using synthetic data as well as using all the TTS data.
%
%The first method, random selection, is commonly considered in the literature. 
%
% For a Neural transducer setup, the confidence score for each word can be approximated by the joiner score of the wordpieces making up the word
Confidence score represents the joiner score of the word-pieces making up the words in an utterance. 
%the probability that the predicted words are correct using the reference as the ground truth. 
We consider the selection of TTS samples with the highest and lowest confidence scores for ASR training.
%The third method computes the 1-best sequence entropy as introduced in \cite{settles2008analysis}.
%The last two methods have been widely considered for data selection in the NLP study. 
%We consider both top and bottom scored utterances as the candidate of selected data for ASR training, in order to figure out the whether the similarity or dissimilarity from the real speech is more important in achieving a better ASR result.   
We compare our methods with these baseline methods on the test sets of Librispeech in terms of word error rate (WER). 

%\subsection{Unit Language Model Based Scoring}
%\label{ssec:lms}
%This method solves the data selection task by using an acoustic unit language model to score an utterance. 

The scoring method using an acoustic unit language model (ULM) is inspired by \cite{lu2022unsupervised}.
The acoustic units are obtained by a Hidden-Unit BERT (HuBERT) model \cite{hsu2021hubert} which is a self-supervised speech model. To emulate a written language having a finite vocabulary of discrete units, a HuBERT model produces the discrete speech representation by directly processing a speech waveform. 
The pseudo labels for HuBERT training are created  by performing unsupervised clustering, such as K-means, on the MFCCs of the input signal. Hence, the learnt discrete speech representation is optimised to be close to the centroid of its belonging cluster, and far from other cluster centroids. It has been shown that fine-tuning a HuBERT model for ASR can reach state-of-the-art results \cite{yang21c_interspeech}. 
In this work, we deploy a HuBERT model trained using K-means clustering method with $K = 100$. %The speech representation of an utterance is fed into a language model for scoring. 
%The resulting speech representation is then modeled with a neural LM to get utterance-level acoustic scores for data selection. 
% \subsubsection{Language model}
% \textcolor{red}{Introduction of language model}
A 13-layer neural LM is trained to predict the next HuBERT unit for a given sequence of input HuBERT units. Please note that this LM is not a conventional text-based LM but one which is trained on discrete speech units and hence, it has the ability to model the acoustic properties. In order to obtain a score per utterance, we used either the average next unit prediction accuracy or the perplexity.

\begin{table*}[ht!]
  \begin{center}
  \caption{Informativeness Comparison in three scoring range using our binary classifier trained with cross-entropy or Arcface loss.} 
  \vspace{-0.1cm}
  \begin{tabular}{l | l | l | c c c c }
    \toprule
    \textbf{Method} & \textbf{Scoring range} & \textbf{\# add. utter. [M]} & \textbf{dev-clean} & \textbf{test-clean} & \textbf{dev-other} & \textbf{test-other}\\
    \midrule
     & \textbf{$Score>0.5$} & \multirow{2}{*}{$0.77$ ($18\,\%$)} & $3.06$ & $3.20$ & $8.95$ & $8.58$\\
    \textbf{Binary classifier} & \textbf{$0.2<Score<0.5$} &  & $2.85$ & $3.11$ & $8.67$ & $8.17$\\
    \cline{2-7}
    \textbf{(Xent)} & \textbf{$Score < 0.2$} & \multirow{2}{*}{$1.10$ ($27\,\%$)} & $2.80$ & $3.13$ & $8.48$ & $8.42$\\
    & \textbf{$0.2<Score<0.5$} &  & $2.78$ & $2.96$ & $8.42$ & $8.16$\\
    \hline
    & \textbf{$Similarity>0.8$} & \multirow{2}{*}{$1.22$ ($30\,\%$)} & $2.85$ & $3.04$ & $8.60$ & $8.34$\\
    \textbf{Binary classifier} & \textbf{$0.2<Similarity<0.8$} &  & $2.81$ & $2.97$ & $8.46$ & $8.09$\\
    \cline{2-7}
    \textbf{(Arcface)}& \textbf{$Similarity < 0.2$} & \multirow{2}{*}{$0.77$ ($18\,\%$)} & $2.96$ & $3.25$ & $9.03$ & $8.45$\\
    & \textbf{$0.2<Similarity<0.8$} &  & $2.91$ & $3.00$ & $8.39$ & $8.19$\\
    \bottomrule
  \end{tabular}
  \label{tab:midwin}
  \end{center}
  \vspace{-0.5cm}
\end{table*}

\begin{table*}[ht!]
\setlength{\tabcolsep}{3pt}
  \begin{center}
  \caption{Number of words existing in Librispeech evaluation set, but not in training set, in terms of different scoring range from binary classifier trained with cross-entropy loss. Number of TTS sentences that contain these words.} 
  \vspace{-0.1cm}
  \hspace{-0.3cm}
  \begin{tabular}{l | c | c c c c c c c c c}
    \toprule
    \multirow{2}{*}{\textbf{Scoring range}} & \textbf{\# add.} & \multicolumn{4}{c}{\textbf{Words}} & &  \multicolumn{4}{c}{\textbf{Utterances}} \\
    \cmidrule(lr){3-6}
    \cmidrule(lr){8-11}
    & \textbf{utter. [M]} & \textbf{dev-clean} & \textbf{test-clean} & \textbf{dev-other} & \textbf{test-other} & & \textbf{dev-clean} & \textbf{test-clean} & \textbf{dev-other} & \textbf{test-other}\\
    \midrule
    \textbf{$Score>0.5$} & $0.77$ ($18\,\%$) & $48$ & $46$ & $49$ & $66$ & & $290$ & $322$ & $234$ & $378$\\
    \textbf{$0.2<Score<0.5$} & $2.24$ ($55\,\%$) &  $157$ & $151$ & $156$ & $182$ & &  $1073$ & $1191$ & $927$ & $1285$\\
    \textbf{$Score < 0.2$} & $1.10$ ($27\,\%$)  & $133$ & $108$ & $118$ & $149$ & & $674$ & $711$ & $514$ & $824$\\
    \bottomrule
  \end{tabular}
  \label{tab:word}
  \end{center}
  \vspace{-0.5cm}
\end{table*}

\begin{table}[ht!]
\scriptsize
\setlength{\tabcolsep}{5pt}
  \begin{center}
  \caption{Testing results, WER [$\%$], of classification scoring methods on Librispeech test set, after removing the LM TTS utterances containing unseen words.}
  \vspace{-0.2cm}
  \begin{tabular}{l | c | c c c c }
    \toprule
    \multirow{2}{*}{\textbf{Method}} & \textbf{Unseen} & \multirow{2}{*}{\textbf{dev-clean}} & \multirow{2}{*}{\textbf{test-clean}} & \multirow{2}{*}{\textbf{dev-other}} & \multirow{2}{*}{\textbf{test-other}}\\
    & \textbf{words} & & & & \\
    \midrule
    \textbf{Bin. cls. } & \XSolidBrush & $2.90$ & $3.09$ & $8.74$ & $8.30$\\
    \textbf{(Xent)} & \CheckmarkBold & $2.78$ & $2.96$ & $8.42$ & $8.16$\\
    \hline
    \textbf{Bin. cls} & \XSolidBrush & $2.88$ & $3.07$ & $8.65$ & $8.20$\\
    \textbf{(Arcface)} & \CheckmarkBold & $2.81$ & $2.97$ & $8.46$ & $8.09$\\
    \bottomrule
  \end{tabular}
  \label{tab:unseen}
  \end{center}
  \vspace{-0.5cm}
\end{table}

\begin{table}[ht!]
\scriptsize
\setlength{\tabcolsep}{2pt}
  \begin{center}
  \caption{Testing results, WER [$\%$], by using the TTS data selected by both ULM and our data selection approaches.}
  \vspace{-0.2cm}
  \begin{tabular}{l | c | c c c c }
    \toprule
    \multirow{2}{*}{\textbf{Method}} & \textbf{\# add.} & \multirow{2}{*}{\textbf{dev-clean}} & \multirow{2}{*}{\textbf{test-clean}} & \multirow{2}{*}{\textbf{dev-other}} & \multirow{2}{*}{\textbf{test-other}}\\
     & \textbf{utter. [M]} & & & & \\
    \midrule
    \textbf{ULM accuracy} & $0.72$ & \multirow{2}{*}{$2.90$} & \multirow{2}{*}{$3.02$} & \multirow{2}{*}{$8.81$} & \multirow{2}{*}{$8.23$}\\
    \textbf{\& Bin. cls. (Xent)} & ($18\,\%$) &  &  & & \\
    \hline
    \textbf{ULM accuracy} & $0.71$ & \multirow{2}{*}{$2.89$} & \multirow{2}{*}{$3.05$} & \multirow{2}{*}{$8.83$} & \multirow{2}{*}{$8.26$}\\
    \textbf{\& Bin. cls. (Arcface)} & ($17\,\%$) &  &  & &  \\
    \hline
    \textbf{LM accuracy} & \multirow{2}{*}{$0.65$} & \multirow{3}{*}{$2.89$} & \multirow{3}{*}{$3.06$} & \multirow{3}{*}{$8.67$} & \multirow{3}{*}{$8.28$}\\
    \textbf{\& Bin. cls. (Xent)} & \multirow{2}{*}{($16\,\%$)} &  &  & &  \\
    \textbf{\& Bin. cls. (Arcface)} &  &  &  & &  \\
    \bottomrule
  \end{tabular}
  \label{tab:fusion}
  \end{center}
  \vspace{-0.7cm}
\end{table}

\subsection{Results}
\label{ssec:res}
%We first train the ASR model by augmenting Librispeech training set with all our generated TTS data, $4.11$ millions of utterances based on LM texts. Compared with the results without using this augmentation, the WERs are reduced from $3.5$ to $2.97$ for test-clean set and from $8.7$ to $8.05$ for test-other set, respectively (\Cref{tab:result}). The data selection methods target at achieving similar WER reduction while using less TTS data.
% Basic result
The ASR model trained using the original Librispeech training set achieves a WER of $3.5\,\%$ on the test-clean split and $8.7\,\%$ on the test-other split (\Cref{tab:result}).
Adding all of our generated TTS data, which consists of $4.11$ million synthetic speech files, as additional training material decreases the WER results on the evaluation sets to $2.97\,\%$ and $8.05\,\%$, respectively. 
We assess our scoring methods for selecting a subset of the TTS audio files for ASR training while aiming to keep the same effectiveness of WER reduction.
With the allowance of $2\,\%$ rising rate, we expect the resulting WER on test-clean to be lower than $3.03\,\%$, and on the test-other to be lower than $8.24\,\%$   

% Baseline result
As seen in \Cref{tab:result}, the data selection methods based on ULM can produce superior results than the other two baseline methods, namely random selection and confidence score-based scoring.
Using the average next unit prediction perplexity as the criterion for data selection, $30\,\%$ of the synthetic audio files with the lowest scores yields a WER of $3.01\,\%$ on the test-clean set and $8.20\,\%$ on the test-other set. To attain the same level of performance, however, we need to incorportate $1.5$ million TTS samples with the highest accuracy scores  when choosing the data based on the average prediction accuracy. In particular, the trained ASR can reach the same WER on the test-clean set as using all of the TTS data available, and a WER of $8.22$ on the test-other split.

%Regarding the baseline methods, the best WER results are $3.05$ and $8.27$ on the test-clean and test-other sets, separately achieved by using additional TTS data selected from the bottom $30\,\%$ confidence scores, and top $30\,\%$ 1-best sentence entropy. Both methods are more effective than random selection. Comparing our proposed solutions to the baselines, we can get better WER results using a single method for both test-clean and test-other at the same time. 

Instead of choosing the highest or lowest scored samples, as was the case with baseline methods, we discovered that our binary classification-based scoring approaches perform best when employing the TTS data within a medium scoring range. In other words, the synthetic audio files that can be differentiated from real speech samples are more beneficial for improving the ASR performance. 
After training our GRU scoring model using BCE loss, we acquire an unweighted average recall of $92\,\%$ classification accuracy on the Librispeech validation set. In particular, the recall score reaches approximately $100\,\%$ for real speech (class 1), and $83\,\%$ for synthetic speech (class 0), demonstrating that almost all the real speech samples can be identified accurately while some of the TTS samples are incorrectly recognised as real speech.  
The scoring model's prediction output are then used to score the TTS samples created from LM text resources. 
The WER can be reduced to $2.96\,\%$ and $8.16\,\%$ for the test-clean and test-other sets, respectively, by choosing only $27\,\%$ of the synthetic audio files rated between 0.2 and 0.5. 
Similar to this, when using Arcface loss to train the GRU model, we can successfully select TTS data that yields an even better result on the test-other set, a WER of $8.09\,\%$, by selecting $30\,\%$ of all TTS audio files which have similarity scores within the range of 0.2 to 0.8.  

\section{Discussion}
\label{sec:discuss}
\vspace{-0.1cm}
To support our recommended scoring ranges for TTS data selection, i.e., score between $0.2$ and $0.5$ for the GRU scoring model trained with BCE loss, similarity between $0.2$ to $0.8$ similarity for the scoring model trained with Arcface loss, we compare their effectiveness to the data scored outside the range.
%To verify the effectiveness of our suggested scoring range in \Cref{ssec:res}, i.e., $0.2$ and $0.5$ for the GRU scoring model trained with BCE loss, and $0.2$ to $0.8$ similarity for the scoring model trained with Arcface loss.
Specifically, for the first scoring model, we randomly select synthetic samples from those scored between $0.2$ and $0.5$ as the same size of the samples scored greater than 0.5, and compare their effectiveness in ASR improvement. Similar comparison has been made for the TTS samples with scores under $0.2$. The random selection was performed $5$ times for each comparison, and the averaged outcomes, shown in \Cref{tab:midwin}, reveal that the TTS samples from our suggested scoring range are more effective in ASR improvement compared to the top- and bottom-scoring samples.
%thanCreftab0.5.%For the case of using BCE loss, we compare the importance of samples that are incorrectly classified as real speech (score greater than 0.5) with the same size of samples randomly selected within the scoring range of $0.2$ and $0.5$. The random selection was run $5$ times, and the average results are given in \Cref{tab:midwin}. It is evident that the TTS samples scored between $0.2$ and $0.5$ are more valuable to the samples that are classified as real speech, which indicates their high similarity. 
However, the samples scored below $0.2$ are also not very helpful for ASR, as shown in \Cref{tab:midwin}. 
This lower-score boundary has been established experimentally; however, its determination can be challenging in practice. Thus, it is suggested to simply choose the highest-scoring TTS samples right below $0.5$.
%The determination of this lower boundary is challenging in practice. We find a proper value by conducting several trials in this work. For convenience, it is suggested to directly employ the top-scoring TTS samples below $0.5
Similarly, for the second scoring model, the TTS samples with too high or too low similarity to real speech are not as competitive as the samples with similarity between $0.2$ and $0.8$ in terms of ASR improvement. Overall, we are able to optimize the ASR performance using TTS samples while excluding samples that are too similar or too dissimilar to real speech.
%Similarly, for the binary classifier trained with Arcface loss, the TTS samples of too high and too low similarity to real speech are not that useful than the samples in the middle. In our experiment, we found out that selecting TTS data of the similarity between $0.2$ and $0.8$ are suitable. Overall, we select data by excluding some improper samples—those too similar to or too different from real speech. 

The results are partly due to the inclusion of out-of-vocabulary words in LM texts like \Cref{tab:midwin}, supporting the idea that we should select TTS data with enough uncertainty as additional ASR training material. 
%
%areof thewords thatCreftab%In \Cref{tab:word}, we list the number of these unseen words exist in the evaluation splits of Librispeech dataset, but not in its training set, and these words are added by LM texts. The TTS samples that are scored lower is partly due to these unseen words; however, including these words in training can help to reduce the uncertainty of the ASR model while decoding. 
%
%When we replace the selected utterances containing these unseen words with other utterances scored within $0.2$ and $0.5$, the results are shown in \ref{tab:unseen}, where the speech recognition performance in terms of WERs becomes worse for both classification scoring methods. 

These results are partially attributable to the incorporation of some extra unseen words given by LM text resources, and as a result, the selected data provide more additional information to benefit ASR training. 
In \Cref{tab:word}, we list the number of out-of-vocabulary words that are included in the evaluation splits of the Librispeech dataset and added to the new training set due to LM texts. An utterance containing these unseen words may have a lower similarity to the original training samples, but including them in ASR training can help to improve speech recognition accuracy.
If the selected utterances containing these unseen words are filtered out for other utterances without them in the same scoring range, the results provided in Table \ref{tab:unseen} show that the speech recognition performance for both of our scoring methods declines to some extent.

%samples,millionCreftabmethods%Additionally, we consider the combination of our classification-based and the ULM scoring method to further suppress the required TTS data size. We train ASR models using only the data selected by both methods, and the best-performing combinations are given in \Cref{tab:fusion}. The LM accuracy and binary classifier (Xent) requires only $0.72$ millions of TTS samples which is about 18.5 percent of the total generated TTS dataset, while it retains the WER results of $3.02 and $8.23 on the test-clean and test-other sets. However, due to the rise of WER on the development side, it is reasonable to suspect the reliability of this ASR model in practical use. 
In an effort to further reduce the required TTS data size, we additionally investigate the combination of our classification-based scoring method with the ULM scoring method . Only the data chosen by both approaches is used to train ASR models, and the best-performing combination is given in \Cref{tab:fusion}. Combining ULM accuracy and a binary classifier (Xent) retains the WER results of $3.02$ and $8.23$ on the test-clean and test-other sets while only requiring $0.72$ million of TTS samples, or about $18\,\%$ of the total generated TTS data. 
However, the WER rise on the development set raises the question of the validity of this ASR model in practical use.

\section{Conclusions}
\vspace{-0.2cm}
In this paper, we presented a solution for selecting useful samples from a given large TTS dataset to augment the ASR training data. Our method uses a simple neural network to produce agreement between real and synthetic speech. Experimental results indicated that the ASR benefits from choosing synthetic data that are of sufficient additional information or uncertainty, resulting in better speech recognition performance compared to other existing methods that consider the maximum or minimum amount of agreement.

\bibliographystyle{IEEEtran}
\bibliography{mybib}

\end{document}